\documentclass[preprint2,final]{aastex}

\usepackage{graphicx}

\newcommand{\el}{$e^-$}
\newcommand{\C}{\arcdeg C}

\slugcomment{DRAFT}

\shorttitle{System description of HAT}
\shortauthors{Bakos et al.}

\begin{document}
\title{System description and first light-curves of HAT, an autonomous
observatory for variability search}

\author{G.~\'A.~Bakos\altaffilmark{1,2}}
\affil{Harvard-Smithsonian Center for Astrophysics\\60 Garden street,
Cambridge, MA02138}
\email{gbakos@cfa.harvard.edu}

\author{J.~L\'az\'ar, I.~Papp, P.~S\'ari}
\affil{Hungarian Astronomical Association, H-1461 Budapest, P.O.Box 219}
\email{jlazar,ipapp, psari@mcse.hu}

\and
\author{E.~M.~Green}
\affil{Steward Observatory, University of Arizona, Tucson, AZ 85721}
\email{bgreen@as.arizona.edu}


\altaffiltext{1}{Predoctoral Fellow, Smithsonian Astrophysical Observatory}
\altaffiltext{2}{Konkoly Observatory, Budapest, H-1525, P.O. Box 67}

\begin{abstract}
Having been operational at Kitt Peak for more than a year, the
prototype of the Hungarian Automated Telescope (HAT-1) has been used
for all-sky variability search of the northern hemisphere. The small
autonomous observatory is recording brightness of stars in the range of
$\rm I_c\approx 6\--13^m$ with a telephoto lens and its
$9\arcdeg\times9\arcdeg$ field of view (FOV), yielding a data rate of
$\sim10^6$ photometric measurements per night. We give brief hardware
and software description of the system, controlled by a single PC
running RealTime Linux operating system (OS). We overview site-specific
details, and quantify the astrometric and photometric capabilities of
HAT. As a demonstration of system performance we give a sample of 60
short period variables in a single selected field, all bright, with
$\rm I < 13^m$, of which only 14 were known before. Depending on the
observing strategy, search for extrasolar planet transits is also a
feasible observing program. We conclude with a short discussion on
future directions. Further information can be found at the HAT
homepage: {\tt http://www-cfa.harvard.edu/\~{ }gbakos/HAT/}.

\end{abstract}


\section{Introduction}

%
The idea of automating ground based observational astronomy goes back
more than two decades. Minimizing manpower can assure uniform and
massive data flow with low budget and the absence of human mistakes.

%

The increasing number of robotic telescopes\footnote{
See eg.~http://alpha.uni-sw.gwdg.de/\~{ }hessman/MONET} 
(capable of computer controlled multiple observations) have been used
for a broad range of projects, such as astrometry:
Carlsberg Meridian Telescope \citep{carlsberg}; 
photoelectric photometry of pre-selected targets: 
Fairborn Observatory \citep{Fairborn}; 
supernova search: KAIT \citep{KAIT}; 
GRB follow-up: ROTSE \citep{Akerlof00,ROTSEIII} 
and LOTIS \citep{Lotis,SuperLotis}, 
exo-planet searches: STARE \citep{STARE}, 
Vulcan \citep{Vulcan}; 
and asteroid searches: TAOS \citep{TAOS}. 
The variety of targets is usually narrow, looking only for
specific timescales, light-curve shapes and intensity ranges.

%
Although most projects concentrating on special targets gain a huge
amount of photometric data, only few of them are capable of presenting
their by-products to the astronomical community, e.g.~OGLE
\citep[e.g.~][]{Wozniak02}, MACHO \citep{Allsman01} and ROTSE
\citep{Akerlof00}.

%
Initiated by ideas of \citet{BP97}, the All-Sky Automated Survey's
\citep[ASAS;][]{GP97} approach is different, in that the final goal
has been photometric monitoring of {\em all bright stars} in a major
part of the southern sky down to $\rm I\approx 14^{m}$. Using a fully
automated but inexpensive system consisting of an amateur-class CCD, a
small telephoto lens and an equatorial mount, ASAS presented catalogues
of 4000 bright variables from a $300 \sq\arcdeg$ area of the southern
sky, {\em 96\% being new discoveries} \citep{GP98,GP00}. The upgraded
ASAS-3 will produce an order of magnitude increase in the data
flow.\footnote{http://www.astrouw.edu.pl/\~{ }gp/asas/asas\_asas3.html}
The incompleteness of our knowledge on bright variable stars was
reinforced by \citet{Akerlof00} who discovered 1781 new variables in a
$2000\sq\arcdeg$ area.

%
Why is general variability study of bright objects important? Several
answers can be found in \citet{BP97,BP00}, and others can be added.
Variable stars are essential for testing stellar structure and
evolution theories, examining galactic structure or establishing the
extragalactic distance scale. Only bright variables are within the
range of high resolution spectroscopy, parallax and proper motion
measurements. Our knowledge of issues related to variable stars
(e.g.~distance scale) can be refined by the combination of detailed
study of close-by, bright objects and of equidistant, homogeneous
samples (e.g.~OGLE \--- Galactic bulge, LMC, SMC). Serious
incompleteness at the bright end affects all conclusions. A systematic,
well-calibrated survey presents clean, statistically valuable samples
with well defined limits for different subtypes of variable objects. A
reliable database with sufficient and ever-growing time-span of
light-curves can be used as an archive, for e.g., correlating optical
variability with X-ray observations, made by satellites. It can be a
valuable input to schedule big-telescope and space-mission
observations, where telescope time is limited, or prior and longer-term
data on field variables is necessary (e.g.~the Kepler mission),
furthermore, it can provide them with a real-time alert system of rare
events. Such events can be nova explosions, helium flash of a star
\citep[Sakurai's object:][]{sakurai1,sakurai2}, 
super-outbursts of dwarf-novae \citep[WZ Sge:][]{wzsge1}.

\begin{figure}[!h]
\epsscale{1.0}
\plotone{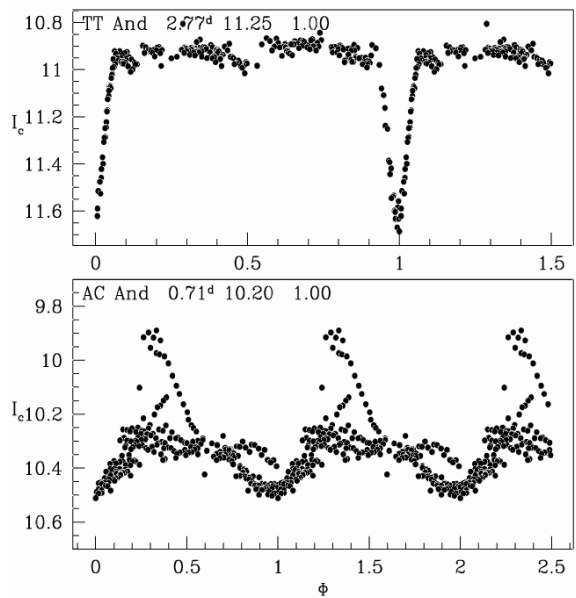}
\caption{Phased light-curves of Algol-type, semi-detached eclipsing
binary TT And, and triple-mode pulsator AC And from HAT observations
(See \S \ref{sec:photprec}). ``Chaotic'' appearance of the AC
light-curve is due to its triple-mode behavior. Phase was computed from
its longest period.
\label{fig:twovar}}
\end{figure}

%
To mention specific examples, observational data is scarce for spotted
red subgiant variables (RS Cvn, FK Com), which are crucial in
understanding the stellar magnetic cycles. Detached eclipsing binaries
(through their stellar mass, radius and luminosity determination) can
be perfect distance and age indicators, if nearby systems are properly
calibrated. Samples of such objects in the solar neighborhood are
sparse \citep{BP97}, partly due to their short and narrow eclipses, and
lack of observational data (see Fig.~\ref{fig:twovar} for our
light-curve of a {\em semi}-detached binary). Bright contact binaries
exhibit similar incompleteness, although long-term observations could
reveal interesting phenomena, such as the transition to semi-detached
state.

The Hipparcos Space Astrometry Mission presented us with discovery of a
few thousand new bright variables, and the Hertzsprung-Russell (HR)
diagram was described in terms of luminosity stability at the
millimagnitude level \citep{hipvarcmd}. However, Hipparcos observed
only selected stars (120000 or $3/\sq\arcdeg$), and the variable star
sample is further limited by the $\sim 110$ epochs per star on average
and cut-off at $\rm I\approx9^m$.

Mapping the location of the large variety of pulsating variables on the
HR diagram is still far from being complete. Addressing phenomena, such
as long-term period and amplitude modulations (e.g.~Blazhko effect of
RR Lyrae), evolution of the pulsational status of a star, is possible
only by long-term and {\em homogeneous} observations. Given the huge
data-flow, interesting phenomena are expected to emerge, for instance
further observational evidence for chaos in W Vir and RV Tau stars
\citep{Buchler95}, triple-mode variables \citep[GSC 40181807:][]{triplegsc}, 
Cepheids which stop pulsating \citep[V19 in M33:][]{Macri01} or strong 
amplitude modulation of Cepheids \citep[V473 Lyr:][]{Burki86}.
Long-term monitoring of semi-regular and Mira variables is needed to
disentangle multiperiodicity and systematic amplitude variations
\citep[e.g.~][]{kiss00}. The sample of the recently established
$\gamma$ Dor subtype (oscillations in non-radial gravity-mode) consists
of only a few dozen stars. One example of the possibilities is our
HAT-1 light-curve of the triple-mode pulsator AC And (Fig.~\ref{fig:twovar}).

%
While the number of robotic telescopes  is around a hundred, there are
only a few completely autonomous observatories, where the human
supervision is eliminated, and all auxiliary appliances (dome, weather
station, etc.) are under a reliable computer control. These
observatories can be installed to remote sites with adequate
astro-climate and infrastructure (electricity, Internet) without need
of an on-site observer, and daily maintenance. The station can be
monitored via Internet, and operation is not a bottleneck any more.

%
HAT is such a small autonomous observatory intended to carry out a
northern counterpart of ASAS, i.e., a variability study of the northern
sky. HAT was developed and constructed by one professional and three
amateur astronomers\footnote{
G.~Bakos (astronomical considerations and software),
J.~L\'az\'ar (software development; www.xperts.hu), 
I.~Papp (electronic design) and 
P.~S\'ari (mechanical engineering)} 
in Hungary, and has been fully operational since May, 2001 at Steward
Observatory, Kitt Peak, Arizona. HAT is controlled by a single Linux PC
without human supervision.

%
The 180mm focal length and 65mm aperture of the telephoto lens, and the
$\rm 2K\times2K$ CCD yields a wide FOV: $9\arcdeg\times9\arcdeg$ on
the sky. Our typical exposure times allow us to study variability of the
$\sim20000$ objects per field brighter than $\rm I_c\approx13^m$ with few
percent precision ($\rm 0.01^m\--0.05^m$), and few minute to one year
time-resolution. As an extreme of the possible observing tactics, HAT
is capable of recording the brightness of {\em every locally and
seasonally} visible star in the range of $\rm I_c\approx 6\--13^m$ in
every second day.
%
%
Our limiting magnitude and photometric precision (See
\S\ref{sec:photprec}) corresponds to the following detection {\em
cut-offs} for a few selected variability types (ranges indicate that
the distance limit depends on the luminosity within the type):
$\gamma$ Doradus stars ($\rm \langle700pc\rangle$), 
$\delta$ Scuti ($\rm 1\--2.5kpc$), 
RR Lyrae ($\rm 3kpc$), Cepheids ($\rm 10\--150kpc$), Miras and semiregular
variables ($\rm \langle60kpc\rangle$).  The limits are based on the
luminosity of the sources, and \--- especially at the larger values \---
overestimate true detection cut-offs, as they do not take into account
reddening and crowding.

With the above specifications, HAT is also suitable for exo-planet
search via transits, which is also included in our program. However, we
concentrate on a broader range of variabilities: a large fraction of
the sky is monitored sparsely, and few selected fields frequently so as
to have preliminary results on short-period changes.

We expect that our survey will contribute to most of the aforementioned
issues related to variability search (public archive, alert system),
and supplement the incomplete variable classes by new discoveries.

Our current data rate is roughly $10^6$ photometric measurement, or two
Gigabytes of raw data per night. Typically a few percent of the sources
are variable, i.e., variable light-curves are supplemented by $\sim
20000$ data points each night.

HAT is monitoring only a fraction of the northern sky, but given the
fact that it is an off-the-shelf system there is perspective for
installation of new units in the near future.

The paper is arranged as follows: \S \ref{sec:hardware} gives an overview on
the hardware, \S \ref{sec:software} describes our software environment, \S
\ref{sec:astro} quantifies the pointing precision of HAT, \S
\ref{sec:installation} and \S \ref{sec:obs} summarize our site-specific
installation at Kitt Peak and observations in the past one year, \S
\ref{sec:photprec} estimates our photometric precision, \S \ref{sec:summary}
gives summary and future directions. 

\placefigure{fig:twovar}
\notetoeditor{It would be important to place Fig.~\ref{fig:twovar}
in the introduction part, so the reader sees some interesting
results before struggling through the techincal part}

\section{Hardware System}\label{sec:hardware}

The HAT hardware consists of an equatorial telescope mount, enclosure
(dome), CCD, a telephoto lens and a PC. Several devices are attached to
the dome and telescope, such as rain-detector, photosensor,
lens-heating and domeflat lights. The PC is protected from inclement
weather by a close-by heated room (``warmroom''). Two parallel port
cables with lightning protection connect the dome to the PC, and are
responsible for driving the telescope and dome devices. Cables for the
CCD depend on the specific setup; in our case a serial line for a Meade
Pictor CCD (for testing purposes) and a custom data cable-pair for an
Apogee AP10 CCD run from the PC to the telescope. A cable in a separate
conduit carries 110V AC to the dome.


\subsection{The Robotic Mount}\label{sec:mount}

Requirements of a mount for massive variability search are: reliability
(for $10^5\--10^6$ actuation), pointing to an accuracy of few orders of
magnitude smaller than the FOV, ability to recover from awkward
positions, never losing orientation, and relatively quick slew-time.

Our mount started as a replica of Gregorz Pojma\'nski's instrument for
all-sky monitoring \citep{GP97}, who kindly shared his plans with our
group. The main mechanical concept of the mount is left unchanged,
i.e., it is a backlash-free friction drive of a horseshoe structure.
Most of the details and dimensions, however, were re-designed. The
mount is machined from forged aluminum alloy, to minimize the
possibility of slow warpage, which could cause uncertainties in the
pointing. All parts are anodized, to yield a tough and resistant
surface.

\begin{figure}[h]
\epsscale{1.0}
\plotone{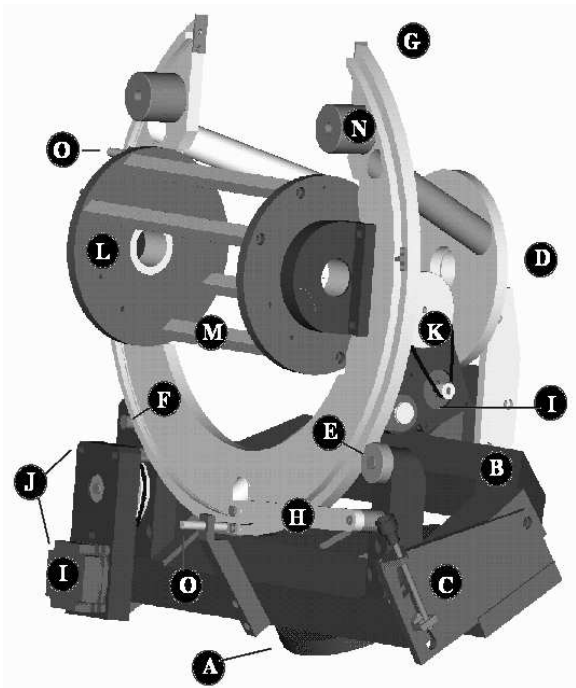}
\caption{HAT mount drawing without the instrument mounting plate, CCD 
and lens. Labels on the image: 
A: main azimuth disc, 
B: rectangular base frame, 
C: altitude groove and adjusting spindle, 
D: RA bearing-housing and polar telescope hole, 
E: RA loose roller, 
F: RA friction-drive, 
G: horseshoe, 
H: horseshoe holding arm, 
I: RA and Dec stepper motors, 
J: RA driving mechanism: sprockets and cogwheels, 
K: Dec driving mechanism, 
L: Dec discs, 
M: Dec lateral bars, 
N: counterweights, 
O: proxy sensors.
\label{fig:mount}}
\end{figure}

The base of HAT is a $\rm \sim200mm \oslash$ (diameter) disc, which can
be mounted on a pier (see Fig.~\ref{fig:mount} for details of the
mount). A rectangular steel plate can smoothly rotate on it, and can be
fine-tuned by two adjusting screws, thus enabling the azimuth setting
during polar axis adjustments.

The base frame of the telescope is held to the steel plate by a fixed
screw and another screw freely running in altitude in a groove, for
both sides. It can be gently adjusted in altitude $\pm15\arcdeg$
(default center is $45\arcdeg$) with a spindle, much like the azimuth
setting, and can be secured with the freely running screw. Inclinations
outside these limits can be easily set by tilting the mounting of the
base plate.

The horseshoe is supported at three points: the bearing-housing, a
loose roller, and the right ascension friction-drive. The horseshoe
outer diameter is $\rm \sim500mm$ with three arms attached to it in
$120\arcdeg$ spacings, forming a cone with pitch angle of
$\sim60\arcdeg$. The apex of this cone is the bearing housing of the RA
axis with a high-precision ball-bearing inside. As the horseshoe is cut
out from a larger piece of plate, relaxation of stress can change its
shape, which affects tracking of astronomical objects. Thus, the whole
structure is machined on a precision lathe as a piece later.

The horseshoe is constrained to the rollers by strong springs holding
an arm and a small bearing running in the circular groove in the inner
side of the horseshoe. The tight contact is crucial for proper
positioning, as the not perfectly balanced horseshoe is friction
driven, thus \--- especially at the extreme positions \--- torque is
needed even for constant motion. The right ascension is driven by a
five-phase stepper motor via sprockets and cogwheels and a
stainless-steel roller with combined gear-ratio of $1$ microstep of the
motor $=1\arcsec$ (arcsec) movement of the horseshoe. Slipping is
further minimized by acceleration and deceleration of the mount in
$\sim50\--100$ discrete velocity steps in an interval (``ramping''),
both specified by the user at the software control. The steel wires of
the sprockets prevent them from stretching, and yield a backlash-free
drive. The maximal slew speed is $\rm \sim2\arcdeg/sec$.

The declination axis is made up of two $\rm 200mm\oslash$ discs
attached by bearings to the inner side of the horseshoe, and connected
by four lateral bars, two of which hold the instrument-mounting plate.
The axis is driven in a similar manner to the RA, i.e., by a 5-phase
stepper, sprocket and cogwheels, but the final resolution is $\rm
5\arcsec/microstep$ and maximal speed is $\rm \sim5\arcdeg/sec$.

The maximal dimension or a telescope which fits our mount is
$\rm\sim200mm\oslash \times 400mm(l)$. 
The maximal size of the detector is limited by the inner half-sphere of
the horseshoe, but most CCDs, such as Apogee AP10
($\rm\sim200\times200\times60mm$), fit readily. Balance of axes is
achieved by counterweights on the top of the horseshoe, on the
declination disc and on the instrument-mounting plate. The telescope
(telephoto lens) is fixed to the instrument-mounting plate by a
converter ring, and rigidly held by screws from the upper two lateral
bars connecting the Dec discs, while the CCD is attached to the other
side.

Inductive proximity sensors on both axes detect home and end positions
in order to ensure fail-safe operation. Even though we use an open-loop
control system (no costly encoder employed), and it might happen that
the telescope loses orientation (due to e.g., power-failure), it can
quickly recover by an iterative procedure of finding the home position.
Physical end positions on both the RA and Dec axes guarantee that even
if the proximity sensors fail, we cannot run off the rollers, and the
telescope tube or detector never hits the mount.

Polar setting of the mount is simplified by the possibility of fitting
a polar telescope to the central hole in the RA bearing house, and
crude setting can be achieved in a few minutes. Final setting is done
by standard methods such as described in \citet{Leung62} and references
therein.

\notetoeditor{Please try to place Fig.~\ref{fig:mount} as close as possible
to \S\ref{sec:mount}}
\placefigure{fig:mount}

\subsection{The dome}\label{sec:dome}

Automated, remote dome operations have to be perfectly fail-safe,
because the detector can be damaged by inclement weather conditions, or
the Sun might edge slowly across the field. A few of the encountered
risks when closing the dome include gusty wind, power outage and
failure of control software. In order to minimize software control and
construction costs, and to enable rapid observation of targets of
opportunity, only mechanisms that completely flip out of the way were
considered. We designed a structure which proved to work in the past
one year, with only a few minor failures. These failures, in turn,
aided improvement of the design.

We experimented with simple schemes, such as a lid opening towards the
north around the upper edge of the box, but this setup has strong wind
resistance during the opening/closing phase. The standard rollover-roof
structure is not compact enough and blocks a considerable part of the
sky.

\begin{figure}[h]
\epsscale{1.0}
\plotone{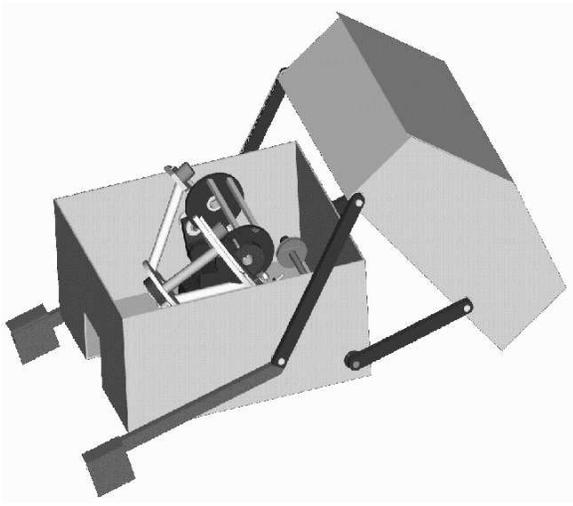}
\caption{Half-open dome of HAT. The asymmetric clamshell structure ensures
fail-safe operation and minimal resistance against wind. 
\label{fig:dome}}
\end{figure}

HAT mount is enclosed by a
$\rm\sim0.6(w)\times0.75(l)\times0.75m(h)$ weather-proof box (See
Fig.~\ref{fig:dome} for a schematic drawing). A separate slant roof on
top is connected to the rest of the dome by swivel joints; a long bar
is attached to its south corner and a shorter bar to the north, on both
sides (assuming an installation on the northern hemisphere). The roof
can be opened northwards by rotation around the bottom axes of the two
arms, resembling an asymmetric clamshell. As the lengths of the arms
differ, the roof co-rotates in such a way that resistance against wind
is minimal in every position, and eventually it gets blocked from the
wind behind the bulk of the dome, with its hollow part facing down. The
longer bar has counterweights on the bottom such that the dome closes
even if the driving mechanism is broken, simply through gravity.

The shorter bar is fixed to a shaft inside the dome, which is driven by
a windshield-wiper DC motor through low gear ratio. Two sensors detect
if the dome reached any of its end positions (closed or open). Safety
power supply for the motor is a 12V 7.5Ah battery, continuously
recharged from the main PC power supply of the dome, which in turn is
connected to the UPS in the warmroom. Even without recharging, this
battery would be able to open/close the dome several times a day for a
week, though there is no need for this, as power failures are detected
by the software, and the system is automatically shut down. Although
normally the PC controls dome operations, a light sensor on the bottom
of the dome acts as a ``big-brother'', and closes the dome if
illuminance reaches that of a dawn/dusk sky. Similarly, a research
grade Vaisala rain-detector has the capability of emergency closing.

Due to problems with skyflat frames (see later in \S\ref{sec:obs}) a
domeflat screen is fixed to the inner north roof, and lit by small
bulbs from positions which yield almost even illuminance not affected
by the telescope's position. When the telescope is pointed to the north
pole, the screen is perpendicular to its aperture. Rotation of the
entire mount about the pole averages out approx.~half of the residual
illuminance pattern.

Dome operations can be also performed manually, overriding other
control. All electronics are installed in a separate compartment on the
south wall, and can be easily dismounted from the outside for servicing
without opening the dome.

\placefigure{fig:dome}
\notetoeditor{Please try to place Fig.~\ref{fig:dome} as close as possible
to \S\ref{sec:dome}}

\subsection{Electronics}
The electronic cards and devices in the dome are connected to the 12V
DC outputs of a simple PC power supply, which in turn gets power from
the warmroom's UPS. Some devices are also connected to the
re-chargeable battery, and can be operated even during power failure.
One card is responsible for control of the mount, another for all other
devices, including the dome.

Signals coming from the parallel port of the PC are optically isolated,
and micro-controller units convert them for direct control of the
stepper motors. Cards are simply {\em converters}, and the
motordrive/clockdrive is the PC's central processing unit, controlled
by the RTLinux OS. 

\subsection{The CCD}\label{sec:ccd}

HAT has been operated with two CCDs. The camera during the test period
of 2000/2001 from Budapest, Konkoly Observatory, was an amateur-class
{\bf Meade Pictor 416xt} camera with a Kodak KAF-0401E chip of
$\rm 512\times768,\;9\mu m$ pixels. This camera yielded noisy images with bad
electronic interference pattern and functioned erratically, but was suitable
for test-mode operation \citep[for more details]{GP97,GB01}. All astrometric
calibration described in \S\ref{sec:astro} was performed with this CCD.

The current Kitt Peak setup uses an {\bf Apogee AP10} camera with
Thomson THX 7899M $\rm 2K\times2K,\; 14\mu m$, grade 2 chip. The chip
covers $\rm \sim 29\times29 mm$, i.e., comparable to the size of small
format photographic films ($\rm 24\times36mm$). Full-well capacity is
$200,000$\el, factory gain setting is $\rm 10e^-/ADU$, thus given the
14-bit (16000 ADU) dynamic range, saturation is slightly limited by the
A/D conversion. Quantum efficiency of the Thomson chip peaks at 40\%
between $\rm 650nm\--800nm$.

Readout noise is variable, being $\gtrsim20$\el.  Bias level is $\rm
270\pm5 ADU$ (nightly variation), with considerable long-term
(day-to-day) instability of $\rm \pm15 ADU$.  Dark current at
$-15$\C \ is $\rm \sim0.05ADU/sec$ and highly dependent on temperature
setting ($\rm \sim0.02ADU/sec$ at $-25$\C). We would like to keep the
temperature of the system at a constant level throughout the year for the
uniformity of the data. Unfortunately the two-stage Peltier cooling is
capable of only $\rm \Delta T\approx 30$\C, and the lowest value we can
achieve is $-15\C$.

The camera is connected to the PC by a data and a control cable plugged
into an ISA-card. The cable length in our setup is 18m, much longer
than the factory default (8m), although well below the company-claimed
maximum limit. This caused problems during the installation, and
hindered us from using the camera for the first 3 months, as the images
contained only a few bits. Typical readout time of a frame is $\rm
\lesssim 10sec$ at 1.3MHz speed, so there is negligible dead-time due
to readout.

Bias frames have a distinct, relatively constant pattern; few bad
columns, many warm pixels, horizontal streaks and clusters of warm
pixels. Dark frames have similar structure, which does not completely
disappear after bias correction. About 10\% of the frames have
anomalous noise and background, the latter is $\sim10\--20$ times
higher than the normal. Most likely these frames are due to a bug in
the readout electronics.

\subsection{The Telephoto Lens}

In principle any kind of telephoto lens or small telescope can be
attached to HAT's mounting plate, which has dimensions smaller than as
described in \S\ref{sec:mount}. Our choice of a Nikon 180mm f/2.8
manual focus telephoto lens ($\rm 64mm\oslash$) for the Kitt Peak setup
was motivated by several factors, such as our limited budget, the
acceptable quality of Nikon lenses, and approximate compatibility with
the ASAS-2 project, which uses 200mm focal length with the same Apogee
CCD \citep{GP02}.

The resulting FOV is $9\degr\times9\degr$, where one pixel corresponds to
$\sim16\arcsec$. The lens gives moderately sharp profiles throughout the
entire field, with half-width of the psf $\sim1.6\--2.0$ pixel
($25\arcsec\--32\arcsec$). The corners show some coma and astigmatism. There
is substantial field-dependent vignetting, reaching a 40\% intensity loss
near the edges, which is not surprising at lenses designed for small-format
photography.

A strong resistive heating of the lens is installed in the light-baffle
(dew cap). This not only prevents formation of dew on the front lens,
but also creates turbulence to blur the stellar profiles to some
extent. As our psf is undersampled, we have deliberately blurred the
image in this way in an attempt to improve photometry \citep[and
\S\ref{sec:photprec}]{GP02}. Slight defocusing of the lens is not
possible, as it introduces strong, spatially dependent distortion of
the profiles.

We observe through a single Cousins I-band filter (Bessel-made), which
has favorable combined sensitivity with the Thomson chip, yielding
considerably higher signal-to-noise ratio for typical stars and
integration times than e.g. a V filter would. The brightness of the
night sky is more stable in I-band at different lunar phases
\citep{Walker1987},
$I \approx 19.9^m/{\sq\arcsec}$ (new moon) to 
$I \approx 19.2^m/{\sq\arcsec}$ (full-moon), as compared to 
e.g.~$V\approx21.8^m,\;20.0^m$, which ensures more uniform photometry. 
New filters can be inserted only manually.


\subsection{The PC}

Any reasonable ($\rm \ga 300MHz$) personal computer which can run a
Linux Operating System and real-time kernels\footnote{As of writing,
Real-Time Linux 3.1, kernels (core of the OS) 2.2.19 or 2.4.$\ast$.},
and has two parallel ports, is suitable for controlling the HAT mount
and dome. An ISA-slot is requisite for controlling the Apogee AP10 CCD,
and serial line is needed for the Meade Pictor CCD.

We use a custom-built PC with 900MHz Athlon AMD processor, 256Mb RAM,
DAT-DDS3 tape archiving facility, and a single 80Mb disc for temporary
storage. Outgoing cables are lightning-protected. Power is from a UPS
plugged in the same circuit as the UPS for the dome power supply. A
watchdog-card automatically hard-resets the computer if that freezes
(does not respond via a special software) for more than 10 minutes.

\section{Software System}\label{sec:software}

Our control PC runs an ordinary Linux OS with a Real-Time Linux
kernel\footnote{www.fsmlabs.com}(denoted RTLinux). Accurate positioning
(e.g.~sidereal speed tracking) needs precisely scheduled stepping
instructions from the control electronics, with the possibility of
frequency tuning (adjusting the tracking speed) and switching between
speeds (e.g.~ramping). These are either directly emitted by a complex
external hardware with built-in frequency standard, {\em or by the
computer} and its central processing unit (in fact a very complex
hardware, but off-the-shelf). Both approaches have pros and cons, but
we found the latter more flexible, because software can be adjusted
remotely, and because it requires less hardware development. The CPU's
capability of emitting signals with a tight schedule depends on the OS.
With ``single task'' operating systems, such as DOS, proper scheduling
of a {\em single} task was possible. However, a further demand is that
simultaneously with signal-generation, one should be able to execute
other tasks, such as CCD operation, disc IO, etc., each of them
requiring some CPU time, thus distorting the frequency of the high
priority processes requesting periodic operations.

RTLinux is such a {\em real-time, multitask} operating system, where
the kernel treats the real-time processes as independent ``threads''
with response times better than $\rm 15\mu s$. The Linux environment,
the user's interface, is run at the lowest priority, but on relatively
fast PCs and assuming only a few real-time processes with no extreme
rescheduling frequencies, there is no noticeable difference to the
user.

\begin{figure}[h]
\epsscale{1.0}
\plotone{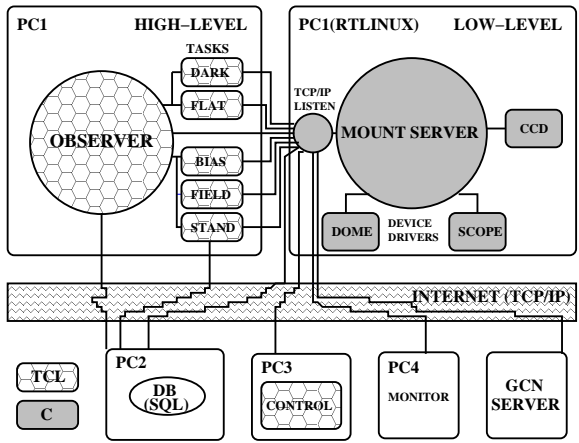}
\caption{Overview of the software system: low and high level control,
central database, remote control and monitoring via Internet. The
``PC'' inscriptions show a possible setup; the computers on which the
given software run, although it is possible to run everything on a
single PC. The programming language is indicated by the backgrounds.
\label{fig:software}}
\end{figure}

The software environment consists of low level programs: {\em scope and dome
drivers, cameraserver, HAT access module} (HAM), and high level software:
{\em central database} (DB) and {\em virtual observer} (``Observer'').
See Fig.~\ref{fig:software} for a schematic flowchart. 

\placefigure{fig:software}

\subsection{Low level programs}
The telescope mount's kernel driver is mainly responsible for
positioning the mount. It is a character device driver, i.e., complex
commands can be issued to it as ascii strings, which are executed after
being parsed.  Status information is readable from a special file,
showing relative positions in steps, RA and Dec, and so on.  Both axes
can independently track (with any sidereal rate), fine-slew or slew
(maximal speed).

As mentioned in \S\ref{sec:mount}, acceleration and deceleration of the
axes is done in finite angular-velocity, or in other words frequency
steps, so as to minimize slip of the friction drive. The ``infinite''
accelerations between the finite velocity levels are in fact smoothed
to continuity by the elasticity of the components. The driver also
reads interrupts coming from the proximity sensors indicating the home
or end positions. It can ``home'' the mount (move marked points on the
horseshoe and declination discs precisely above the proximity sensors)
from any unknown position, which is handy for resetting the
continuously accumulating slips, and for orientation if the telescope
is ``lost''. The home position's hour angle and declination is
calibrated at the installation of the telescope. After nulling values
at the home position, the number of steps taken in any axis and any
direction is always stored, and the residuals after a subsequent homing
indicate the amount of slip. The tracking speed can be fine-tuned with
a precision of $10^{-7}$, which means that the hardware element's
absolute dimensions (e.g., the ratio of the driving roller to the
diameter of the horseshoe) are not that restricted (but only important
in absolute pointing).  Finally, RA and Dec values can be assigned to
any position.

The dome driver is a similar RTLinux kernel driver responsible for i)
turning the main power on/off, ii) closing or opening the dome, iii)
activating lens heating, iv) control of domeflat lights, v) detection
of rain. Status of the dome can be queried from the driver.

CCD acquisition software of the Apogee AP10 CCD is based on the
application programming interface from \citet{GP02}, modified to
fulfill our needs. The camera server can run as a standalone
application, or as a ``daemon'', accepting commands from other software
via Internet Protocol (TCP/IP).

The so-called HAT Access Module (HAM) is the main low-level program
joining all the aforementioned codes. It accepts input via TCP/IP from
higher level software (e.g., the virtual Observer), parses the
commands, and distributes them to the relevant resources. That is, all
telescope, dome and CCD operations are piped through this program.
Another aspect of HAM is allocation of the drivers, i.e. no other
program can issue conflicting commands to the hardware.

\subsection{High level programs}
We use the fast, reliable and open source relational Structured Query
Language (MySQL\footnote{Information on MySQL is available at
www.mysql.org}) database for storing associated parameters of the
station (setup of mount, telescope, CCD, scheduled tasks, etc.), and
for keeping continuously accumulating logs of operations (parameters of
archived images, etc.), all of them arranged into tables. This has
several advantages over simply keeping information in files without a
database wrapper: i) faster operation on data, ii) TCP/IP access (for
instance through the Internet), iii) use of existing interfaces to
MySQL from various programming languages. The smooth TCP/IP
communication enables installation of multiple HAT observatories in any
preferred topology, with central database management, and with the
possibility that the sites interact with each other, for example,
request the last object observed or instruct another telescope.

The observatory as a whole is managed by the {\em virtual} Observer,
which communicates with the access module, central database and
executes (observer) programs. This is a Tcl\footnote{Information on Tcl
is available at www.scriptics.com} language-based simulation of
near-perfect observer, modeled as a finite-state machine, which makes
decisions/transitions depending on the circumstances.

%
The Observer programs (such as taking flatfield frames and all-sky
monitoring) are run as independent threads, so the Observer never hangs
by waiting for a time-consuming operation to finish; it can be aborted
any time. Tasks have start times (relative to sunset), allowed
durations and priorities \--- both taken from the DB, whose properties
are used by a task-manager (part of the Observer) to launch or stop
them and to suspend or interrupt other tasks. There is a template,
which makes writing new tasks very simple for the user, using a command
library of a few hundred commands, such as ``ObserveObject''.

%
During normal operation, the Observer is in ``run'' state. If the
weather is unfavorable, the Observer passes to a ``weather sleep''
state, where the CCD is kept cooled, the dome is closed, and the system
waits for clearing. In the daytime the system is in ``daysleep'', power
is turned off, CCD is warmed back, and we are waiting for the first
scheduled task to appear within the characteristic time needed for
starting up. If the Internet connection with the monitoring stations is
lost, system jumps to ``suspend'' state, turns off power, and waits a
long time before finally exiting to ``service'' state. Recovery from
service is possible only by manually restarting the system.

%
Upon entering the ``run'' state, Observer checks the tasks to be
executed during the given session, checks weather information, and if
needed, starts up the system (turns power on, starts cooling the CCD).
Following this, it enters an endless cycle, and breaks out only in case
of transition to another state. During the cycle it responds to direct
TCP/IP messages from outside, and checks if connection with the access
module and outside-connection monitor is alive. The ephemerides are
continuously updated, such as apparent position of the Sun and Moon.
The task manager launches, checks or stops tasks, administrates changes
in the DB. If a task exits abnormally, perhaps due to some kind of
exception (programming error), it is flagged, and not launched again
any more. Proper scheduling of tasks is one of most complex part in the
code.

The weather status is checked every $\rm \sim10 sec$, and if
needed, the current task is aborted, and the observatory transits to
another state. There is a possibility of looking for targets of
opportunity by parsing remote TCP/IP packets, such as the GRB
Coordinates Network (GCN)\footnote{http://gcn.gsfc.nasa.gov/}. The
Observer checks and can take action upon incoming email communication,
which enables the scientist to control or monitor the observatory even
from a cellphone.

Eventually, with a given time resolution ($\rm \sim30min$), 
ConCam\footnote{For more information: www.concam.net} \citep{Pereira00} 
all-sky night-time images and infrared satellite images\footnote{
National Weather Service \--- www.wrh.noaa.gov} 
are downloaded and stored. These proved to be very useful later in
flagging the photometric quality of a night. The Observer also has the
ability to recover from various emergency situations (e.g., disk is
full, device missing), and notify recipients via email.

Automatic start up of HAT is included in the booting procedure of the
computer, similarly, closing the observatory in the shutdown phase.
This way the service staff can easily start up/bring down the system.
Also, HAT operations resume as the power outage ends.

\section{Pointing precision of HAT}\label{sec:astro}

While highly accurate pointing of a telescope covering a 9 degree field
might appear to be merely a luxury, we tried to have a general purpose
design with the flexibility of upgrading the focal length, thus
decreasing the field of view, which requires more precise pointing.
Moreover, sufficient pointing capability of even a wide field
instrument can be relevant to its photometric precision. There are
indications that precise aperture photometry depends on positioning the
same star over the same pixels of the CCD, due to inter and intra-pixel
variability \citep{Buffington91,Robinson95}, although minute
displacements of the telescope also do have advantages in filtering out
internal reflections due to the fast focal ratio of the instrument. Due
to our open-loop system, deviation of absolute pointing accumulates
proportionally to the individual pointing accuracy, which has to be
minimized.

Astrometric calibration was carried out at Konkoly Observatory,
Budapest during the summer of 2001. We used a replica (HAT-2) of the
mount at Kitt Peak, the Meade Pictor 416xt CCD, and a Soviet ``MTO''
$\rm 100mm\oslash$, f/10 Maksutov telephoto lens\footnote{Special
thanks to A.~Holl for lending us the lens.}. The 1m focal length and
the fine, $\rm 1.86\arcsec/pixel$ resolution were useful for
calibration, which used the starry sky as reference grid. Whenever
needed, we made approximate correction for atmospheric refraction, and
used Guide Star Catalogue \citep[GSC;][]{GSC} for finding the
astrometric solution of the frames, and coordinates of the central
pixel.

We distinguish between the following attributes of pointing: {\em
homing, tracking, absolute-pointing} precisions and {\em pointing
repeatability}. In general, depending on the kind of precision, crucial
factors can be the balance of the axes, ramping-up distance, maximal
slewing speed, surface conditions and shape of the horseshoe, polar
setting and orthogonality of the axes (RA, Dec, optical axis).

Homing precision measures how precisely can we set the mount to the
home-position defined by the proximity sensors. The scatter of
individual homings turned out to be $\rm \sigma_{RA}=0.5sec$, $\rm
\sigma_{Dec}=25\arcsec$.

The tracking of the system depends on adjustment of the polar axis,
tracking speed (set by the mount driver), and the effect of
environmental changes (e.g., strong wind) on the axes. The projected
position of the mount's polar axis on the sky was determined from the
arcs of stars from exposures with $\approx 20\times$ sidereal tracking
around the pole, during which the diurnal motion of the stars was
negligible. As actual position of the celestial pole is known from
astrometry databases, we could adjust the polar axis with $\rm \pm
20pix\approx 0.5\arcmin$ precision.

\begin{figure}
\epsscale{1.0}
\plotone{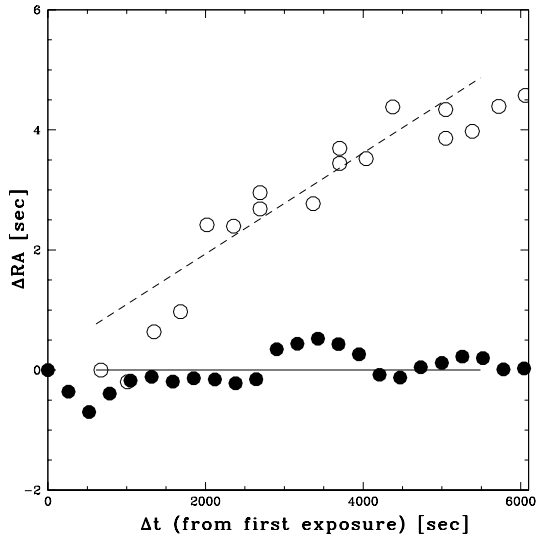}
\caption{Tracking errors of HAT and MTO $\rm 100mm\oslash$ lens in a 2 hour
run. The upper line and empty circles show improperly tuned tracking
speed. The lower line represents calibrated tracking. Deviations with
characteristic time $\rm \sim10^3sec$ arise from the
irregularities on the RA roller. \label{fig:tracking}}
\end{figure}

The tracking was measured in calm weather conditions on fields
culminating near zenith, from hour angle $\rm HA\approx-1h$ to $\rm
HA\approx1h$. Tracking speed was adjusted and then kept fixed in such a
way that tracking residuals were minimal.  The overall tracking error
during 2 hours was less than $\rm 0.5sec$ in RA. We also found
quasi-periodic errors with $\rm\sim20min$ characteristic timescale and
$\rm 0.5 sec$ amplitude, which were probably due to irregularities on
the small RA roller (see Fig.~\ref{fig:tracking}).

\placefigure{fig:tracking}

We defined absolute pointing precision as the precision of moving the
telescope to a celestial object (grid star) with respect to the home
position or another star (reference star). Pointing errors scaled with
the arc of movement, but overall scatter {\em without} removing any
systematic error with a correction map were in the order of
$\rm \sigma_{RA}\approx 20sec$,
$\rm \sigma_{Dec}\approx70\arcsec$. 
Repeatability was measured by moving the telescope back to the
reference point, usually a bright star close to zenith.  Repeatability
is better than absolute pointing:
$\rm \sigma_{RA}\approx7sec$,
$\rm \sigma_{Dec}\approx50\arcsec$. 
This indicates that our absolute pointing errors \--- at least partly
\--- originate from non-perpendicularity of the axes and imperfect
polar axis setting.

\section{Installation of HAT-1 on Kitt Peak}\label{sec:installation}

\begin{figure}
\epsscale{1.0}
\plotone{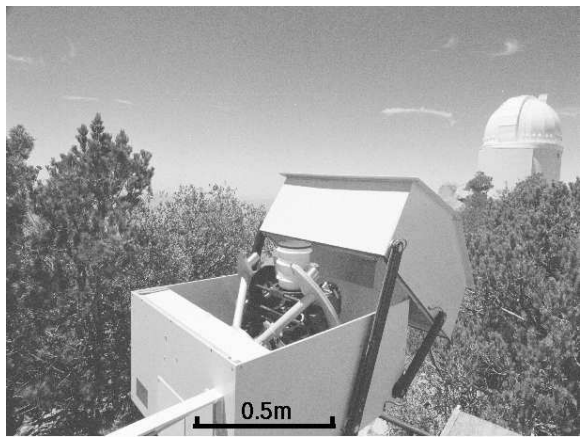}
\caption{
HAT-1 at Kitt Peak: asymmetric clamshell dome and
horseshoe-mount. The white tube is the light baffle on the Nikon lens. 
\label{fig:kpno_setup}}
\end{figure}

%
\placefigure{fig:kpno_setup}

HAT-1 was transported (airmail) within its dome from Budapest to
Steward Observatory, Tucson in January, 2001. The telescope was badly
damaged during the trip, but thanks to Steward Observatory's
hospitality, it was repaired in a few weeks. The telescope was
installed to Steward Observatory, Kitt Peak in the following month, and
was ready for operations by March 2001.

Our PC is hosted by the SuperLotis building's warmroom, and cables go
out to the $\rm \sim2m$ high, massive concrete pillar on the western
edge of the ridge, holding the HAT dome and telescope.  The bug (\S
\ref{sec:ccd}) with the AP10 CCD delayed operation of HAT until
mid-May, but finally Apogee serviced the camera.

The first few month period was devoted to debugging the system and
finalizing observing programs. Although HAT worked in robotic mode, its
operation was monitored from Hungary. As reliable dome status
(open/close) information is not available for any of the domes at Kitt
Peak, we had to set aside the idea of slaving our dome opening to that
of a manually operated telescope, and open/close jointly with the other
dome. Thus, weather status was assessed at early Kitt Peak evening (5am
Central European Time) from the Internet using the daytime webcamera of
Kitt Peak\footnote{http://www.noao.edu/cgi-bin/kpno/axim.cgi}, the
National Weather Service forecast\footnote{http://www.wrh.noaa.gov} and
ConCam images during the night.  Some information is automatically
downloaded and parsed by HAT, e.g.~we switch to ``weather-sleep'' state
if the wind sensors of the 4m Mayall telescope show gusts in excess of
45km/h.

During the debugging period we started regular observations of selected
fields. The logfiles were examined next day, and updated software was
transferred to the site by the next evening. This way, in a few months,
the software became very reliable, and no major modifications have been
performed since.

Since September, 2001, HAT has been operated from the
Harvard-Smithsonian Center for Astrophysics, Cambridge, MA.

\section{Observations from Kitt Peak}\label{sec:obs}

A typical observing session of HAT is described in the following.
Roughly 1.5 hours before sunset, camera cooling is started, and after
the service temperature is reached, 20 bias frames are taken. Following
this, twenty 4-minute dark frames are exposed. If weather is clear, the
``skyflat'' task starts up 10 minutes after sunset. After opening the
dome, HAT selects an optimal flatfield region from the database, which
is close to the preferred point of minimal sky gradient
\citep{Chromey96}. This point is usually opposite to the Sun in
azimuth, and $20\arcdeg$ from zenith. The telescope is randomly moved
between skyflat frames so as to permit median averaging out brighter
stars, and exposure time is recursively tuned to keep the central
intensity at $\sim50$\% of saturation. Even with these considerations,
the twilight sky on a $9\arcdeg\times9\arcdeg$ area has gradients up to
10\%, and therefore we installed a domeflat screen. Unfortunately this
brought up further complications, and dome-flat observations are still
under development. At the end of the night session, skyflats, darks and
biases are taken again.

The monitoring of selected fields (``monfield'') starts when the Sun
sinks $11\arcdeg$ below the horizon. The entire sky was split up into
696 slightly overlapping fields, each $8\arcdeg\times8\arcdeg$ wide.
Choosing the next field to be observed is done by a sophisticated
algorithm. A list of enabled fields is loaded from the database, and
visible ones with high enough elevation are selected. Visibility means
that the object is within the limits of the mount, above the artificial
horizon-grid, and far enough from the Sun and Moon ($\ga 45\arcdeg$).
Ranking of the fields is done by combining their previous observation
times (the more recent, the lower rank), their proximity to the
meridian or the western horizon (depending on the ranking method) and
manually set priorities from the database. After any observation,
parameters (e.g.~time of last observation) are updated and synchronized
with the DB. This ensures that none of the regions are observed too
frequently until there are other, favorably situated fields. Fields
close to culmination, thus having the smallest possible airmass and
differential refraction, are not missed. Priorities of fields can be
such that given their visibility, they are observed with higher
frequency, or even exclusively. Finally, a small random factor is
appended to the ranks in order to avoid the same selection on two
consecutive nights, thus daily aliases in frequency analysis. The
telescope is homed every hour to keep pointing accuracy.

Our present observing tactic is to observe most of the visible fields
sparsely, for information on long-time variation, and only concentrate
on a few fields for short-term variability. The high priority is
re-assigned to different fields on a weekly \-- monthly timescale. With
two consecutive, randomly displaced 240s exposures per field, we can
carry out photometry in the range of $\rm I_c\approx 6-13^m$. We also
experimented with 30s exposures so as not to lose information on very
bright, saturated stars, but this increased the data flow without
considerable amount of extra photometric information.  With this
approach we collect about 80 (winter) to 50 (summer) frame-pairs per
clear night, which means approx.~30 photometric points per night and
star for high priority fields ($10^6$ photometric measurement
altogether).

Calibration of the data to standard $\rm I_c$-band is not
straightforward, as we currently observe only with a single $\rm I_c$
filter. Either Hipparcos Catalogue stars \citep{Leeuwen97} in our
fields are used for direct calibration, or we observe Landolt standards
\citep{Landolt83,Landolt92} throughout the night. In both cases we have
to restrict ourselves to color-independent terms:
\[i_{inst} = I_c + \xi_i + k_i\cdot X -\dot{\xi}\cdot(t-t_0),\] 

where $i_{inst}$ and $I_c$ are  the instrumental and (close to) standard
magnitudes, $X$ is the airmass, $t$ is the time from an arbitrary epoch
$t_0$, and $\xi_i$, $k_i$, $\dot{\xi}$ are coefficients to be
determined by the regression.

Observation of Landolt standards is performed by the ``standard'' task,
which selects suitable standard stars (bright enough, not extreme
colors, have large airmass and hour-angle span), and observes the star
with fine time and airmass resolution plus the culminating fields. This
task is launched maximum few times a month, during
absolute-photometric, new-moon nights.

Current data flow with lossless compression is $\rm \sim1.6Gb/day$
(winter) to $\rm 1Gb/day$ (summer), thus new DAT DDS-3 tapes have to be
inserted every week.

The only unautomated procedures in HAT operation are checking weather
status at evening (rain is detected automatically, but clouds not) and
tape changing.

During a year's operation HAT has completed approx.~$140$ observing
sessions, with a total number of $\sim35000$ exposures of which 19000
were field observations, the rest were calibration frames.

\section{Photometric precision of HAT}\label{sec:photprec}

Reduction of the current 160Gb of data is under way. Photometric
precision of HAT-1 at Kitt Peak was tested using a 12-night subset of
images from September/October 2001 for the moderately crowded field
``F077'' ($\rm \alpha=23^h15^m00^s$,$\rm
\delta=48\arcdeg00\arcmin00\arcsec$).  This contains only about 5\% of
our current data set. We primarily concentrate on the repeatability of
the measurements, i.e., the precision for generating light-curves (in a
system as close to $\rm I_c$ as possible) as compared to the {\em
accuracy}, which is relative to absolute standards. Our crude estimate
for the latter from Hipparcos I-band stars, standard star observations
and the problematic flatfielding is that absolute calibration errors
can be as high $\rm 0.1^m$ in the field corners, but less than $\rm
0.05^m$ in the center.

A summary of the sessions is listed in Table \ref{tab:photsum}.
Observations were taken during a servicing mission, thus calibration
frames are not available for all nights, and the system was not
providing its maximal performance. Baffle on the lens was installed
only in January 2002, and overscan region of the chip was not read out
properly. The field was observed to zenith angles as high as
$60\arcdeg$.

\begin{table}[h]
\caption{Summary of HAT-1 test observations\label{tab:photsum}}
\begin{tabular}{l|c|c|c|c|c|l}\hline
 Date & N & Bias & Dark & Flat & Moon & Com \\\hline\hline
21/09 & 43 & 0  & 0  & 0  & 20\% & Clear\\
22/09 & 62 & 0  & 0  & 0  & 30\% & Clear\\
24/09 & 42 & 0  & 0  & 0  & 50\% & Clear\\
25/09 & 52 & 0  & 14 & 53 & 60\% & Clear\\
26/09 & 38 & 16 & 18 & 0  & 70\% & Clear\\
27/09 & 32 & 4  & 18 & 0  & 80\% & \\
04/10 & 58 & 20 & 18 & 0  & 99\% & \\
05/10 & 62 & 26 & 18 & 34 & 90\% & Cirrus\\
10/10 & 36 & 25 & 18 & 35 & 50\% & Cirrus\\
12/10 & 26 & 19 & 18 & 3  & 20\% & Cirrus\\
13/10 & 26 & 24 & 18 & 34 & 15\% & Cirrus\\
14/10 & 27 & 24 & 18 & 29 & 10\% & Cirrus\\\hline\hline
\end{tabular}
\textsc{Note.\---} The table only shows data used for finding out
the photometric precision of HAT-1, and comprises $\sim 5\%$ of the complete
data set. ``Date'' format is day/month, year 2001. ``N'' denotes the number
of frames for field F077. ``Moon'' is the phase of the Moon.
\end{table}

Calibration of frames was done in a standard way, and consisted of bias
and dark subtraction, and flatfield correction. Time-dependence of bias
level was not dealt with. For some of the nights dark frames were not
available, however the dark pattern seems to change in time.
Calibration was done by our Tcl interface to \textsc{iraf}
\footnote{
IRAF is distributed by the National Optical Astronomy Observatories,
which are operated by the Association of Universities for Research
in Astronomy, Inc., under cooperative agreement with the National
Science Foundation.} \citep{IRAF}. 

As mentioned in \S\ref{sec:obs}, flatfielding with skyflat frames of a
$9\arcdeg$ wide field is problematic. We tried to remove the large scale
pattern (partly due to division with flatfields not truly representing
the response function of the system) by blank-sky correction, using
median average of frames taken at moonless, clear nights, close to
zenith, and far from the Milky Way. Unfortunately even these frames
should be treated with caution because in some cases gradients from
zodiacal light or the rippled structure of the night airglow were
easily visible, and confirmed from ConCam images.

Astrometry of the master frame was performed by
\textsc{iraf/geomap,geoxytran}, and the residuals (offsets in
arcseconds) of the second order polynomial transformation between the
celestial reference frame and the image were in the order of 
$\rm \sigma\approx0.8\arcsec$.

Photometry of images was finally done by the aperture photometry
routine in the \textsc{iraf/daophot} package \citep{DAOPHOT} after
experimenting with \textsc{isis-2.1} Image Subtraction Method
\citep{Alard00} without positive results due to our undersampling
(fwhm=$1.6\--2.0$ pixel). The bottleneck in the latter method seemed to
be spatial interpolation of the images to the reference grid, and
re-sampling the narrow psfs. Experiments with \textsc{iraf/geotran} to
get around this problem did not significantly improve the results.

An astrometric reference image was montaged from ~20 individual frames
by transforming them to the same reference grid. A starlist was
established by \textsc{daofind} (60K stars), and later the same IDs
were assigned to the same stars on all frames for easy
cross-identification. This astrometric reference was useful {\em only}
for finding sources with good efficiency, but due to the resampling of
narrow psfs during the transformation of individual images, it was not
used in photometry at all.

We found roughly 30K stars per individual frame down to a threshold of
5 sigma of the background with \textsc{daofind}. Background scatter
($\rm \sigma_{bg}\approx10\--20ADU$) almost entirely comes from the sky
(mean level $\rm \approx 300ADU$ at new moon) and processing noise
\citep{Newberry91}, while readout noise ($\rm R\approx2ADU$),
digitization noise ($\rm T\approx0.29ADU$), etc., are negligible. The
raw photometric data was first transformed to the master frame in the
XY-plane, stars were cross-identifed, IDs were re-assigned.

A reference magnitude file was created by averaging individual raw
photometry measurements for a dozen good quality images. Magnitudes for
all frames were shifted to this reference, where the amount of the
shift was spatially dependent, typically done on a $6\times4$ XY-grid,
and few hundred stars for each block. The residuals of the
transformation indicated our photometric precision, and were in the
order of $\rm \sigma\approx0.01^m$. Spatial dependence is a reasonable
assumption due to the wide field-of-view and differential extinction.
Some fraction of the scatter is due to color-dependence of extinction,
which we cannot take into account. Standard calibration of the field
was tied to non-saturated Hipparcos stars.

\begin{figure}[h]
\epsscale{1.0}
\plotone{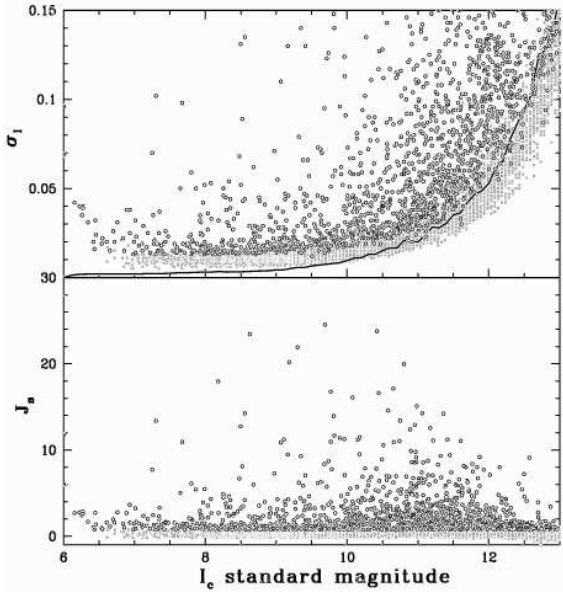}
\caption{Photometric precision of HAT with 8 minute exposures. The
upper panel shows the rms of light-curves, which contain at least 100
points. Open circles denote stars with $\rm J_s$ variability index
greater than $0.75$, i.e., suspected variables, while gray points mark
constant stars. The solid line shows the average {\em estimated} error
by \textsc{daophot}. The lower panel shows the variability index for
the same stars.
\label{fig:photprec}}
\end{figure}

After construction of the light-curves, we could estimate the
photometric precision of the system by assuming that most of the stars
are constant, and by deriving the rms of the light-curves around their
mean values. Most light-curves have 250 data points, and each point is 
the result of $\rm 2\times240s$ exposures. Comparison with the formal
error of individual points given by the photometry code (based upon
the flux of the star, the background, and various parameters) showed
that the formal error underestimates our rms with stars brighter than
$\rm I\sim10^m$, while slightly overestimates at faint sources (see
Fig.~\ref{fig:photprec}). We also derived the $J_s$ Stetson variability
index
%
%
for each star \citep{Stetson96} following \citet{Kaluzny98}, where the
formal errors are scaled by a linear relation to the true errors
(Fig.~\ref{fig:photprec}, lower panel). 

\placefigure{fig:photprec}

Stars with at least 100 data points and $\rm J_s > 0.75$ were selected
as suspected variables ($\sim1700$). After standard Fourier analysis,
the selection was further narrowed by looking only at periods shorter
than $\rm 25^d$ and Signal Detection Efficiency
\citep{Alcock00,Kovacs02} of the main peak in the power spectrum
greater than 4.5. Finally, 60, mostly short period light-curves were
selected by hand (see Fig.~\ref{fig:multivar}). We cross-correlated the
list with the General Catalogue of Variable Stars
\citep[GCVS]{GCVS}, including the New Suspected Variables (NSV), and found
only 12 known and 1 suspected variables. Stars were also looked up in
the Hipparcos \citep{HIP}, Tycho-2 \citep{Tycho2} catalogues, Tycho
variable stars \citep{Piquard01a}, and the Simbad database\footnote{We
acknowledge the use of the SIMBAD database, operated at CDS,
Strasbourg, France: simbad.u-strasbg.fr}, but only one (out of those
not in the GCVS) was found with variability flag (GSC 0362701580 in
Fig.\ref{fig:multivar}). Eight stars out of the suspected new
discoveries were in the well-known BD, HD or SAO catalogues. 

In addition, we retrieved all variables in our FOV from the GCVS
catalogue having brighter maximum than $\rm 13^m$, not being saturated
on our images, and having a period shorter than $\rm 25^d$ (14
entries). Ten out of them were found by our survey, one was too
crowded, and omitted from the original coordinate lists (CZ And),
another variable was too elongated in the corner of our field, thus not
found by \textsc{daofind}. CU And, semi-detached binary was excluded
from our selection due to its small variability index ($\rm J_s =
0.5$), although it has a good quality light-curve. Finally, FL Lac was
faint and crowded, and our light-curve was too noisy for selection.

The 46 new variables are marked with their GSC numbers. Several
long-period variables were also found, but due to the limited time-span
of our observations, we do not include them in the present paper.

\section{Summary and future directions}\label{sec:summary}
We described a small, autonomous observatory, which has been working
for one year at Steward Observatory, Kitt Peak. In spite of the small
telephoto lens used as the ``telescope'', it can perform massive
photometry of bright sources. It completed on the order of $20000$
pointings to different objects, yielding a data flow of $\sim 10^6$
photometric measurements per night, and proved to work reliably. Using
5\% of our data, and a single selected field, a few dozen bright
($I<13$) variables were found, further reinforcing the incompleteness
even on the bright end of previous variability searches. Some
parameters of the system are summarized in Table~\ref{tab:spec}.

HAT-1 is capable of monitoring a {\em small fraction} of the sky with
sufficient time resolution. Multiple filters would not only ease
classification of sources based upon their colors, but due to
differential refraction (vs.~color), the photometric precision would be
also improved.  Wider aperture would increase the incoming flux, and
either our time-resolution at constant photometric precision would
drop, or our limiting magnitude with the same exposure times would be
expanded. This would not be necesseraly {\em improvement} of the
system, but in some sense the target of observations would be
different. HAT is designed to operate in a network. It is an
off-the-shelf system of $\rm \sim15K$ USD cost, plus the CCD. These
make multiple installations easy. It is also a flexible, multi-purpose
observatory, not only eligible for all-sky monitoring, but could be
used as a photometric monitor station at bigger observatories.  Current
bottleneck is neither hardware development or operating HAT, but rather
efficient data reduction, archiving and web-availability of the data.
After efficient software can handle the data flow and overcome the
difficulties, we plan to upgrade HAT with more filters, bigger
telescopes, and to install multiple stations.

\begin{table}[h]
\caption{Specification of HAT-1 at Kitt Peak. 
\label{tab:spec}}
\begin{tabular}{lr}\hline
\multicolumn{2}{c}{Equatorial mount}\\\hline\hline
Max.~RA slew & $\rm 2\arcdeg/$sec\\
Max.~Dec slew & $\rm 5\arcdeg/$sec\\
RA resolution & $\rm 1\arcsec/$step\\
Dec resolution & $\rm 5\arcsec/$step\\
Max.~tel.~diam. & 20cm\\\hline
\multicolumn{2}{c}{CCD (Apogee AP10)}\\\hline\hline
Dimensions & $\rm 2K\times2K$, $\rm 14$ micron\\
Gain & 10 e-/ADU\\
Readout noise & 2 ADU\\
Dark current	&	$\rm 0.05ADU/sec$ at $-15\C$\\
Readout time & $\rm <10s$\\
Cooling & Peltier ($\rm \Delta T=32\C$)\\\hline
\multicolumn{2}{c}{Lens (Nikon 180mm f/2.8)}\\\hline\hline
Aperture & 65mm\\
Plate scale & $\rm 16\arcsec/pixel$\\
Field of view & $9\arcdeg\times9\arcdeg$\\\hline
\multicolumn{2}{c}{Pointing precision}\\\hline\hline
Homing  & $<0.5\arcmin$\\
Tracking & $\rm <0.5sec/2hr$\\
Absolute & RA:$\rm 20sec$, Dec: $1\arcmin$\\
Repeatability & RA: $\rm 7sec$, Dec: $1\arcmin$\\\hline
\multicolumn{2}{c}{Photometry ($\rm I_c$ band, 8min time res.)}\\\hline\hline
Accuracy (absolute) & $\rm \approx 0.05^m$\\
Precision $\rm (I_c < 10^m)$& $\rm \lesssim 0.01^m$\\
$\rm I_c=11^m$ & $\rm 0.02^m$\\
$\rm I_c=12^m$ & $\rm 0.05^m$\\\hline\hline
\end{tabular}
\end{table}

\acknowledgments
\section{Acknowledgements}
This project was initiated by Prof.~Bohdan Paczy\'nski. We are grateful
for his tireless support, encouragement, advice and funds from the
generous gift of Mr.~William Golden. We are indebted to G.~Pojma\'nski
for his collaboration and for sharing his plans and software with us.
It is a pleasure to thank P.~Strittmatter the opportunity of installing
HAT to Kitt Peak, the partial funds in the installation costs and the
hospitality of Steward Observatory. We are thankful to the whole staff
of Steward Observatory (B.~Peterson, G.~Stafford, W.~Wood, J.~Rill) for
their consistent support in operating HAT. G.~Bakos wishes to
acknowledge the kind hospitality of Konkoly Observatory during the test
period of HAT, while being an undergraduate and junior research fellow,
with special thanks to the director, L.~G.~Bal\'azs. The project was
partially funded by the Hungarian OTKA T-038437 grant. Support is also
acknowledged to NASA grant NAG 5-10854.  G.~Bakos is greatly indebted
to the Smithsonian Astrophysical Observatory for support through the
SAO Predoctoral Fellowship program. We thank G.~Kov\'acs, R.~W.~Noyes,
D.~D.~Sasselov, K.~Z~Stanek and A.~H.~Szentgyorgy for valuable
discussions and their careful reading of this manuscript.

\begin{figure}
\includegraphics[bbllx=57,bblly=-31,bburx=554,bbury=712,scale=0.9]{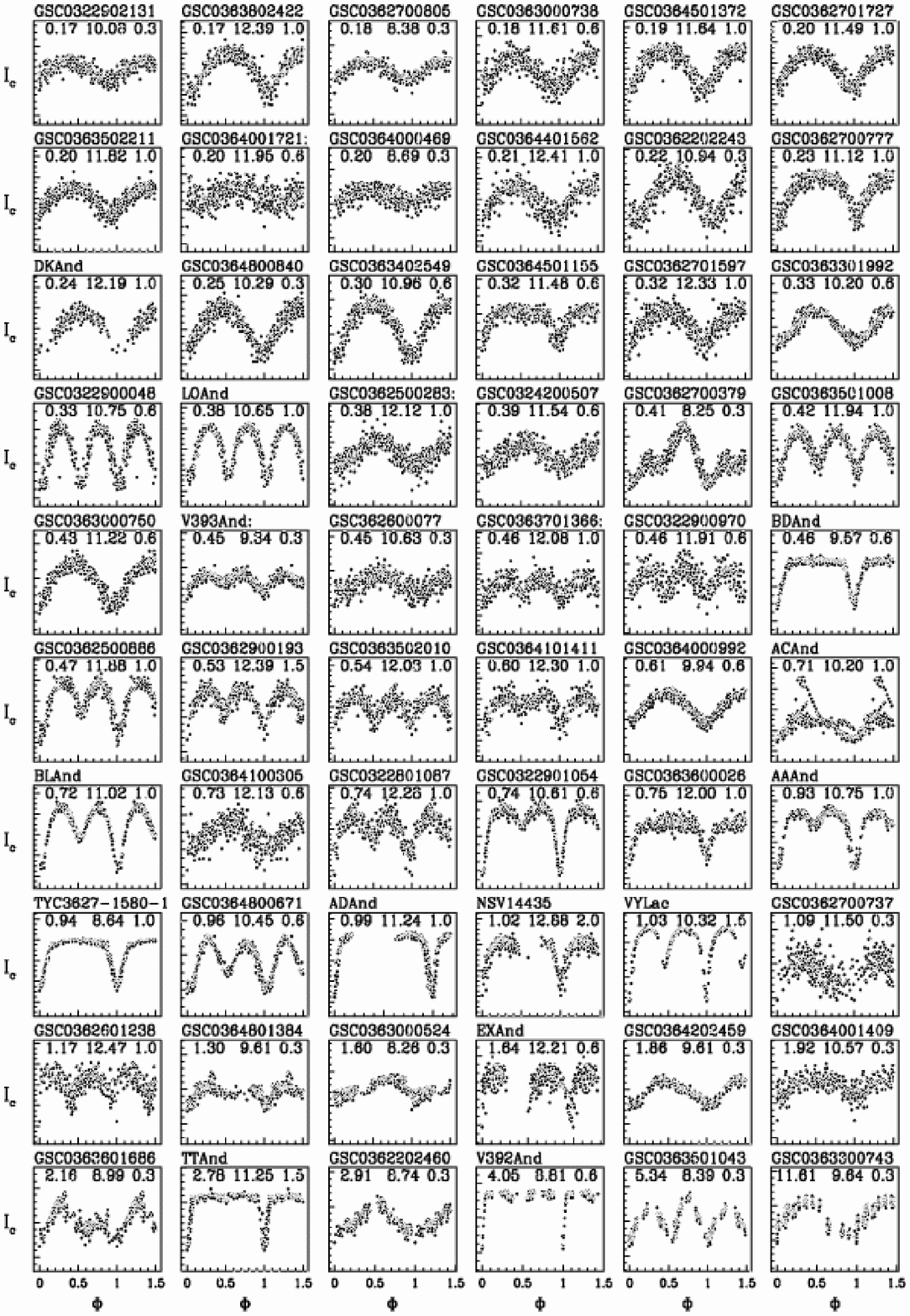}
\vskip-70pt
\caption{Selected variables from HAT observations of field ``F077''. Known
variable stars are marked with their conventional names, new
discoveries with their GSC numbers. Numbers in the panels indicate the
period, average I-band brightness and the height of the box
(magnitudes). Colons (:) denote uncertainty in the
cross-identification. 
\label{fig:multivar}}
\end{figure}

%
\clearpage
\notetoeditor{Figure \ref{fig:multivar} should cover a whole page}
\placefigure{fig:multivar}
\end{document}